\documentclass[conference]{IEEEtran}
\IEEEoverridecommandlockouts

\usepackage{cite}
\usepackage{amsmath,amssymb,amsfonts}
\usepackage{algorithmic}
\usepackage{graphicx}
\usepackage{textcomp}
\usepackage{xcolor}
\usepackage{multirow}
\usepackage{hyperref}
\usepackage{tabularx,booktabs}
\newcolumntype{Y}{>{\centering\arraybackslash}X}

\def\BibTeX{{\rm B\kern-.05em{\sc i\kern-.025em b}\kern-.08em
    T\kern-.1667em\lower.7ex\hbox{E}\kern-.125emX}}
\begin{document}

\title{A Case Study on Modeling Adequacy of a Grid with Subsynchronous Oscillations Involving IBRs \\
\thanks{This study is supported by the Advanced Grid Modeling (AGM) Program of the Office of Electricity, the Department of Energy (DOE).}}

\author{\IEEEauthorblockN{$^1$Lilan Karunaratne, $^1$Nilanjan Ray Chaudhuri,  $^2$Amirthagunaraj Yogarathnam,  and  $^2$Meng Yue}
\IEEEauthorblockA{ \textit{$^1$School of Electrical Engineering and Computer
Science}, Pennsylvania State University, University Park, PA, USA \\
\textit{$^2$Interdisciplinary Science Department}, Brookhaven National Laboratory, Upton, NY, USA\\
emails: lvk5363@psu.edu, nuc88@psu.edu, ayogarath@bnl.gov, yuemeng@bnl.gov}}

\maketitle
\begin{abstract}
A case study on modeling adequacy of a grid in presence of renewable resources based on grid-forming converters (GFCs) is the subject matter of this paper. For this purpose, a $4$-machine $11$-bus IEEE benchmark model is modified by considering GFCs replacing synchronous generators that led to unstable subsynchronous oscillations (SSOs). We aim to: (a) understand if transmission network dynamics should be considered in such cases, (b) revisit the space-phasor-calculus (SPC) in $d$-$q$ frame under balanced condition that captures such phenomena and lends itself to eigenvalue analysis, and (c) emphasize limitations of such models while underscoring their importance for large-scale power system simulations. Time-domain and frequency-domain results from SPC and quasistationary phasor calculus (QPC) models are compared with electromagnetic transient (EMT)-based simulations. It is shown that models with transmission line dynamics in SPC framework can capture the SSO mode while QPC models that neglect these dynamics fail to do so.   
\end{abstract}

\section{Introduction}
The growth of inverter-based resources (IBRs) has introduced unfamiliar dynamics into the traditional power system. At present, power systems employ two primary IBR technologies: (a) grid-following converters (GFLCs) and (b) grid-forming converters (GFCs). Among these, the first one is commonly adopted whereas GFC represents a relatively new technology under grid-connected operation. In the recent past, bulk power systems in different parts of the world have witnessed subsynchronous oscillations (SSOs) involving GFLCs. The IEEE Power \& Energy Society (PES) IBR SSO task force has recently compiled a list of $19$ such events \cite{IBRSSOTF}. Root cause analysis fundamentally divides such phenomena into three classes $-$ (a) series capacitor SSO, (b) weak grid SSO, and (c) inter-IBR SSO. A comprehensive list of literature in this area can be found in \cite{IBRSSOTF}, whereas the possibility of inter-IBR SSO was shown in \cite{Lingling-InterIBR}.

With this background, it is obvious that a clear understanding of the interaction of IBRs with the rest of the components in the grid is required as the IBRs start dominating the generation portfolio. This underscores the importance of studying modeling adequacy of the traditional quasistationary assumption of transmission networks in planning models. More specific, we focus on the modeling adequacy of balanced systems with multiple IBRs that are solely based on GFC technology. 

Literature on modeling adequacy of IBR-dominated systems in the context of SSOs is quite limited. A five-step methodology for determining modeling adequacy of grids with IBRs was presented in \cite{Strunz-21-QPC-DPC}. These steps are -- (1) dynamic modeling, (2) frequency response analysis, (3) modal analysis, (4) sensitivity analysis, and (5) validation through time-domain simulation. The paper shows that quasistationary phasor calculus (QPC)-based models can produce inaccurate results compared to dynamic phasor calculus (DPC)-based models in presence of GFLCs, GFCs, and synchronous generators (SGs). 

Notwithstanding the importance of \cite{Strunz-21-QPC-DPC}, the paper does not clearly articulate the DPC-based modeling framework when it comes to interfacing SG models with stator transients and IBR models, with the lumped parameter dynamic model of the transmission network; given different frames of references needed in the process. The other aspect that requires further clarity is the interchangeable usage of \textit{baseband abc-frame representation} and the \textit{time-varying dominant Fourier coefficient-based representation} as `dynamic phasor' (DP).

The first paper that developed a comprehensive calculus of the baseband abc-frame representation for bulk power system model was \cite{Mani-94-DynamicPhasor}, which is also followed in \cite{Strunz-21-QPC-DPC}. On the other hand, the theory of generalized averaging proposed in \cite{Verghese-91-DynamicPhasor} led to the second representation, which has widely been called DPs in papers including \cite{Stankovic-00-DynamicPhasor}, \cite{Bozhko-16-DynamicPhasor}, and many others. As pointed out in \cite{Mani-94-DynamicPhasor}, it is essential to limit the bandwidth of the phasors to strictly below the carrier frequency ($50$ or $60$ Hz in power systems), which however is not a constraint for a Park's transformation-based approach proposed in \cite{Mani-95-DQ0} that is essentially based on space-phasor-based calculus (SPC) in rotating $d$-$q$ frame. The above discussion shows that some clarity is needed regarding the common features and differences among the multitude of approaches describing the time-varying phasors.

In contrast with the existing literature, in this paper we consider power grids in which all IBRs are GFCs.  The contributions of this paper are the following -- (1) a comparative summary is presented that clarifies different notions of the so-called \textit{time-varying phasors} in literature; (2) a space-phasor-based modeling approach in a rotating $d$-$q$ frame for a generic power system with SGs and droop-controlled GFCs is presented with sufficient clarity; (3) a case study on a modified IEEE $4$-machine test system is performed to develop understanding of the criticality of network dynamics in SPC-bases models compared to algebraic representation in their QPC counterpart; and (4) the advantages and limitations of SPC-based models for planning studies are emphasized. 

\section{Different Representations of Time-Varying Phasor Calculus}\label{sec:TimeVarPhasor}
The notion of time-varying phasor calculus has taken three different forms, which are elaborated below. Please see Chapter $3$ of \cite{Demiray2008} for a review on this topic. 

\textit{1. Baseband-abc representation:} This representation is first proposed in \cite{Mani-94-DynamicPhasor}, where a \textit{modulated single-phase} signal $x(t) = X(t)cos(\omega_s t + \theta(t)) \in \mathcal{B}$ can be mapped into a time-varying phasor of the form $x_{bb}(t) = X(t)e^{j\theta(t)} \in \mathcal{L}$, where $\omega_s$ is the synchronous speed in electrical rad/s. The phasor operator $\Upsilon$ can be defined as a mapping $\Upsilon: \mathcal{B} \rightarrow \mathcal{L}$ such that $x_{bb}(t) = \Upsilon(x(t))$ and $x(t) = \Re{\{x_{bb}(t)e^{j\omega_s t}\}},~\forall t \in \mathbb{R}$, where, $\Re\{\cdot\}$ is the real part operator. The operator $\Upsilon$ is essentially a composition of transforming the modulated signal $x(t)$ to the analytic signal $x_a(t) = x(t) + j\mathcal{H}\{x(t)\}$ and the frequency shift operation leading to $x_{bb}(t) = x_a (t)e^{-j\omega_st}$, where $\mathcal{H}\{\cdot\}$ denotes the Hilbert transform. \vspace{-5pt}

\begin{figure}[h!]
    \centerline{    \includegraphics[width=0.4\textwidth]{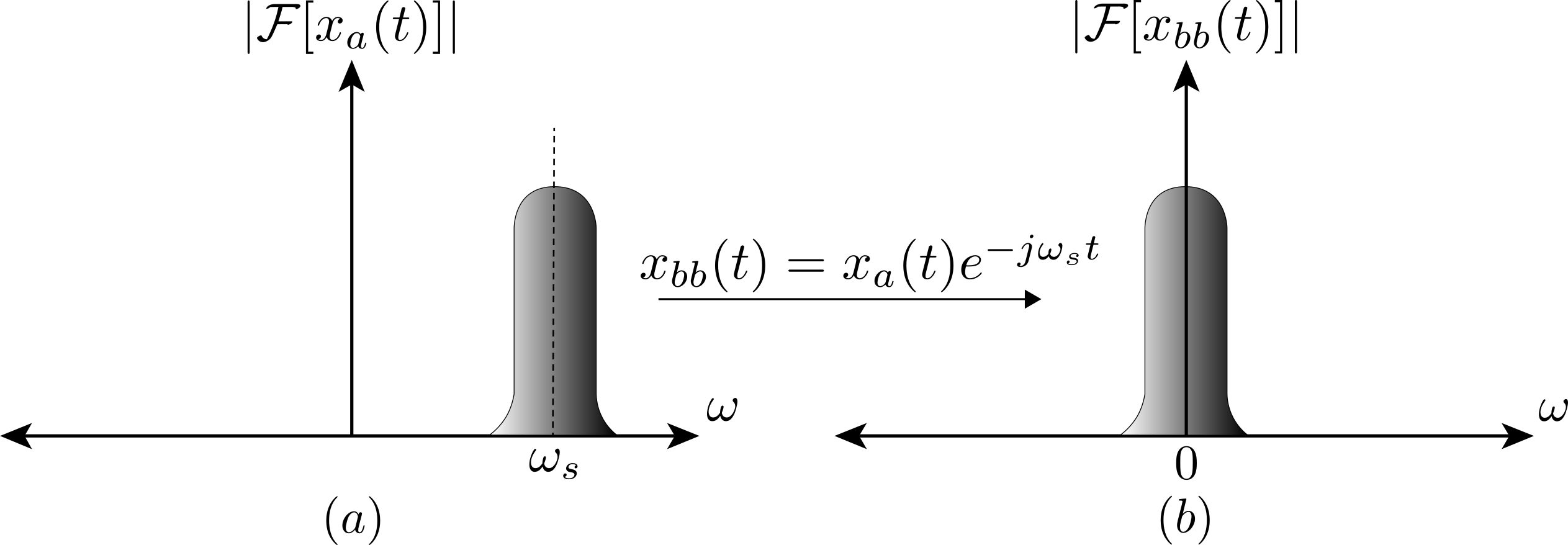}}
    \vspace{-5pt}
    \caption{Signal spectrum of (a) an analytic and (b) a baseband signal.}
    \label{fig:spectrum}
\end{figure} 
\vspace{-5pt}

The following properties of $\Upsilon$ are essential for its practical application in power systems.\\
\textit{$\Upsilon: \mathcal{B} \rightarrow \mathcal{L}$ should be (a) bijective, and (b) linear.}\\
The first property ensures that the phasor operator is well-defined and the second property is essential for satisfying KCL, KVL, and power balance of the network in time-varying phasor domain. 

It was established in \cite{Mani-94-DynamicPhasor} that these properties will hold if $\mathcal{L}$ is restricted to the set of low-pass phasors and as a consequence $\mathcal{B}$ becomes the set of band-pass signals as dictated by $x(t) = \Re{\{x_{bb}(t)e^{j\omega_s t}\}},~\forall t \in \mathbb{R}$. In other words, $\mathcal{F}\{x_{bb}(t)\} = \mathbf{X}_{bb}(j\omega) = 0, \omega \geq \omega_s ~\text{and} ~\omega \leq -\omega_s$ and $\mathcal{F}\{x(t)\} = \mathbf{X}(j\omega) = 0, \omega = 0 ~\text{and} ~\omega \geq 2\omega_s$, where $\mathcal{F}\{\cdot\}$ is the Fourier transformation. These constraints are shown in Fig.~\ref{fig:spectrum}. This representation can be applied to individual phases of the power system for balanced condition as well as unbalanced condition using sequence transformation.

Finally, for $x(t), \frac{\mathrm{d} x(t)}{\mathrm{d} t} \in \mathcal{B}$ the derivative operation satisfies the following relationship \vspace{-2pt}
\begin{equation}
\Upsilon\left ( \frac{\mathrm{d} x(t)}{\mathrm{d} t} \right ) = \frac{\mathrm{d} x_{bb}(t) }{\mathrm{d} t} + j\omega_s x_{bb}(t).
\end{equation}

\textit{2. Space-phasor-based representation in $dq0$ frame:} The space-phasor (also called space-vector) $\bar{x}(t)$ is defined as the transformation of arbitrary three-phase signals $x_a(t)$, $x_b(t)$, and $x_c(t)$ such that $\bar{x}(t) = x_\alpha (t) + jx_\beta (t) = \frac{2}{3}\begin{bmatrix}
1 & \alpha & \alpha^*
\end{bmatrix}\textbf{x}(t)$, where $\alpha = e^{j\frac{2\pi}{3}}$, $(.)^*$ denotes conjugate operation, and $\textbf{x}(t) = [x_a(t)~ x_b(t) ~x_c(t)]^T$. In other words, this transformation leads to \textit{orthogonalization} of the $abc$ frame quantities. If $x_a(t) + x_b(t) + x_c(t) \equiv 0$, then $\bar{x}(t)$ holds the instantaneous information of $\textbf{x}(t)$.

Of particular interest is the special case of \textit{balanced} set of \textit{modulated} signals \cite{Mani-95-DQ0} of the form $x_a(t) = X(t)cos(\omega_s t + \theta(t))$, $x_b(t) = X(t)cos(\omega_s t + \theta(t) - \frac{2\pi}{3})$, and $x_c(t) = X(t)cos(\omega_s t + \theta(t) - \frac{4\pi}{3})$ for which $\bar{x}(t) = X(t)e^{j(\omega_s t + \theta(t))}$. The time-varying phasor $\bar{x}_{bb}(t)$ is calculated by the frequency shift operation on $\bar{x}(t)$ leading to $\bar{x}_{bb}(t) = x_d(t) + jx_q(t) = \bar{x}(t)e^{-j\rho(t)}$, where $\frac{\mathrm{d} \rho (t) }{\mathrm{d} t} = \omega (t)$. The operator $\bar{\Upsilon} : \mathcal{M} \rightarrow \mathcal{D}$ is defined as $\bar{x}_{bb}(t) = \bar{\Upsilon}(\textbf{x}(t))$, where $\mathcal{M}$ is a vector space representing the set of balanced three-phase signals. It can be shown that 
$\bar{\Upsilon}(\textbf{x}(t)) = [1~j~0] \mathcal{P}(t)\textbf{x}(t)$, where $\mathcal{P}(t) = \frac{2}{3}\begin{bmatrix}
\cos\rho(t) & \cos\left ( \rho(t) - \frac{2\pi}{3} \right ) & \cos\left ( \rho(t) + \frac{2\pi}{3} \right ) \\ 
-\sin\rho(t) & -\sin\left ( \rho(t) - \frac{2\pi}{3} \right ) & -\sin\left ( \rho(t) + \frac{2\pi}{3} \right ) \\ 
\frac{1}{2} & \frac{1}{2} & \frac{1}{2}
\end{bmatrix}$ 
is the Park's transformation matrix. 

Notice that $\bar{\Upsilon}(\textbf{x}(t))$ is \textit{a bijective linear transformation} since $\mathcal{P}(t)$ is invertible. Finally, for $\textbf{x}(t) \in \mathcal{M}$ the derivative operation satisfies the following relationship \vspace{-3pt}
\begin{equation}  \vspace{-2pt}
\bar{\Upsilon}\left ( \frac{\mathrm{d} \textbf{x}(t)}{\mathrm{d} t} \right ) = \frac{\mathrm{d} \bar{x}_{bb}(t) }{\mathrm{d} t} + j\omega(t) \bar{x}_{bb}(t).
\end{equation} 
Also, if $\omega(t) = \omega_s$ is chosen as in \cite{Mani-95-DQ0}, then $\bar{x}_{bb}(t) = x_d(t) + jx_q(t) = \bar{x}(t)e^{-j\omega_s t} = X(t)e^{j\theta(t)}$, which is similar to the baseband-$abc$ case. 

\textit{3. Generalized averaging theory-based representation:} In \cite{Verghese-91-DynamicPhasor} the generalized averaging theory was proposed, which expresses a near-periodic (possibly complex) time-domain waveform $x(\tau)$ in the interval $\tau \in (t - T, t]$ using a Fourier series of the form $x(\tau) = \sum_{k = -\infty}^{\infty} X_k(t) e^{jk\omega_s\tau}$, where $\omega_s = \frac{2\pi}{T}$, $k \in \mathbb{Z}$, and $X_k(t)$ are the complex Fourier coefficients that vary with time as the window of width $T$ slides over the signal. The $k$th coefficient, also called the \textit{$k$th phasor}, can be determined at time $t$ by the following \textit{averaging} operation $X_k(t) = \frac{1}{T}\int_{t-T}^{t} x(\tau) e^{-jk\omega_s\tau}d\tau = \left \langle x \right \rangle_k (t)$.

The derivative operation satisfies the following relation
\begin{equation}
    \left \langle \frac{\mathrm{d} x}{\mathrm{d} t} \right \rangle_k (t) = \frac{\mathrm{d} \left \langle x \right \rangle_{k}(t)}{\mathrm{d} t} + jk\omega_s\left \langle x \right \rangle_{k}(t)
\end{equation}
Note that the time-domain waveform $x(\tau)$ above can be $abc$ phase quantities or $dq0$ frame quantities. 

In \cite{Stankovic-02-DP}, the generalized averaging method has been applied to the three-phase case to determine the $k$th dynamical +ve, -ve, and 0-sequence components $\left \langle x \right \rangle_{p,k}(t)$, $\left \langle x \right \rangle_{n,k}(t)$, and $\left \langle x \right \rangle_{z,k}(t)$, respectively, in the following form $\begin{bmatrix}
\left \langle x \right \rangle_{p,k}(t) & \left \langle x \right \rangle_{n,k}(t)  & \left \langle x \right \rangle_{z,k}(t)
\end{bmatrix}^T = \frac{1}{T} \int_{t-T}^{t}e^{jk\omega_s\tau}\mathcal{T}^H 
\textbf{x}(\tau)d\tau$, where, $\mathcal{T} = \frac{1}{\sqrt{3}}\begin{bmatrix}
1 & 1 & 1\\ 
\alpha^* & \alpha & 1\\ 
\alpha & \alpha^* & 1
\end{bmatrix}$ and $(.)^H$ denotes Hermitian operation.

In this approach we are interested in a good approximation provided by the set $\mathcal{U}$ of dominant Fourier coefficients such that $x(\tau) \approx  \sum_{k \in \mathcal{U}} \left \langle x \right \rangle_{k}(t) e^{jk\omega_s\tau}$.

\textbf{\textit{Important remarks}}\\
1. Baseband-$abc$ phasor operator $\Upsilon$ is bijective and linear as long as the time-varying phasor's speed is restricted by the low-pass assumption mentioned earlier. On the contrary, no such restrictions on the speed is required in the space-phasor-based representation in $dq0$ frame.\\
2. The relationship between the space-phasor $\bar{x}(t)$ and the dynamic sequence components \cite{Stankovic-02-DP} is $\bar{x}(t) = \frac{2}{\sqrt{3}} \sum_{k = -\infty}^{\infty} e^{jk\omega_s t} \left \langle x \right \rangle_{p,k}(t)$. As $k \in \mathcal{U}$ is considered, the generalized averaging-based method leads to an approximated model. On the contrary, the space-phasor-based calculus in $dq0$ frame is an accurate representation as it only depends on a transformation.  \\
3. For analyzing unbalanced systems and harmonics, generalized averaging gives significant computational advantage compared to $dq0$ frame models.\\
4. For balanced systems, which is the focus of this paper, we consider the space-phasor-based calculus in $d$-$q$ frame as our framework of choice, which we will refer to as the SPC-based approach from now on. For notational convenience, we will use $\bar{x}_{bb} = x_{dq} = x_d + jx_q$ and drop the time variable $t$.

\section{Modeling in SPC and QPC Frameworks}
This section briefly discusses the detailed mathematical representation used for modeling a generic power system with SGs, GFCs, transformers, transmission lines, and loads.  

\subsection{Modeling of transmission network and loads}\label{sec:TrNw_LoadMdl}
The transmission network in the SPC-based representation uses a lumped $\pi$-section model as shown in Figs \ref{fig:gen_ref_frame} and \ref{fig:circuit_gfc} consisting of the following KCL and KVL algebraic equations.\\ \vspace{-5pt}
\begin{equation}\label{eq:KCL_VL}
  \mathbf{i_{NDQ}} = \mathbf{CCI}\times \left [ \mathbf{i_{DQ}}^T~\mathbf{i_{lDQ}}^T \right ]^T;~\mathbf{v_{lDQ}} = \mathbf{CCU}\times\mathbf{v_{NDQ}}  
\end{equation}
The model uses a synchronously rotating $D$-$Q$ frame. Assuming the network has $n$ nodes, $l$ series $R$-$L$ branches excluding $m$ SG/IBR transformers, $\mathbf{i_{NDQ}} \in \mathbb{C}^n$ is the vector of net injected currents in each node coming from shunt capacitance and any load that may be present, $\mathbf{i_{lDQ}} \in \mathbb{C}^{l}$ and $\mathbf{i_{DQ}} \in \mathbb{C}^{m}$ are the vectors of currents flowing through each series $R-L$ branch and SG/IBR transformer, respectively, $\mathbf{v_{lDQ}}\in \mathbb{C}^l$ is the vector of voltage drops across series $R$-$L$ branches, $\mathbf{v_{NDQ}} \in \mathbb{C}^n$ is the node voltage vector, and $\mathbf{CCI} \in \mathbb{R}^{n\times (l+m)}$ and $\mathbf{CCU} \in \mathbb{R}^{l\times n}$ are the incidence matrix and nodal connectivity matrix, respectively. 

Constant impedance loads are represented using dynamic models of parallel $R_L$-$L_L$-$C_L$ elements. 
The following differential equations describe the transmission line and load model, where $\omega_s$ is the synchronous speed in electrical rad/s and the remaining quantities are in per unit (p.u.).
\begin{equation} \label{eqn:con_dyn_eqn}
    \begin{array}{l}
         \dot i_{lDQ} = \frac{\omega_{s}}{L_{l}}[ v_{lDQ} - j\omega^* L_{l} i_{lDQ} - R_{l} i_{lDQ}] \\
         \dot v_{NDQ} = \frac{\omega_{s}}{(\frac{C_{l}}{2} + C_{L})}[i_{NDQ} - i_{LDQ} -\frac{v_{NDQ}}{R_{L}} \\ \hspace{40pt}- j \omega^* (\frac{C_{l}}{2} + C_{L}) v_{NDQ}] \\
         \dot i_{LDQ} = \frac{\omega_{s}}{L_{L}}[ v_{NDQ} - j\omega^* L_{L} i_{LDQ}] \\
    \end{array}
\end{equation}
Here, $\omega^* = 1.0$ p.u. and $i_{lDQ}$, $i_{NDQ}$, $v_{lDQ}$, and $v_{NDQ}$ are the elements of corresponding vectors in \eqref{eq:KCL_VL}.

The QPC-based model adopts a transmission network which is represented using admittance matrix-based algebraic equations. Standard current injection framework in the synchronously rotating $D$-$Q$ reference frame is utilized to solve for the bus voltages \cite{pai}. The static loads with constant impedance characteristics are assumed in this model.

\subsection{Modeling of SG including stator transients}\label{sec:SG_Transients}
\begin{figure}[h!]
    \centerline{
\includegraphics[width=0.395\textwidth]{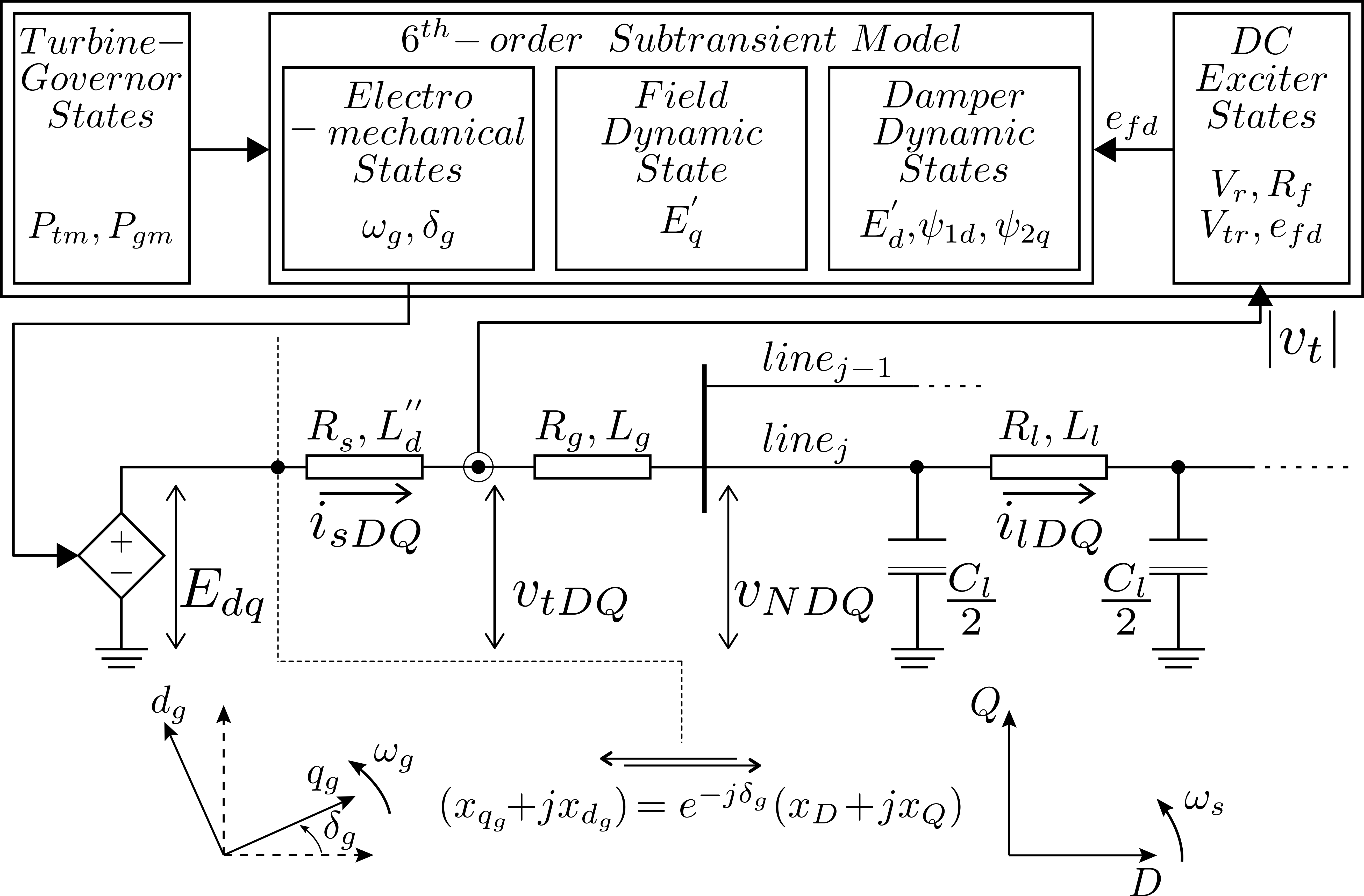}}
    \vspace{-5pt}
    \caption{Interconnection between SG model and transmission network.}
    \label{fig:gen_ref_frame}
\end{figure}

The QPC model of SG considers a $6$th-order subtransient model in its $q_{g}$-$d_{g}$ reference frame rotating at corresponding rotor speed $\omega_g$ along with turbine, governor and exciter dynamics as shown inside a box in Fig.~\ref{fig:gen_ref_frame}. The model assumes a leading $d$-axis per IEEE convention, neglects stator transients and subtransient saliency (i.e., $L_d^{''} \approx L_q^{''}$), and has  $\omega_{g}$, $\delta_{g}$, $E_{d}^{'}$, $E_{q}^{'}$, $\psi_{1d}$, $\psi_{2q}$ as dynamic states, see pp. $99$ of \cite{chaudhuri2014multi}. 

\textit{The SPC model of SG considers stator transients, whose interface with the dynamic model of the transmission systems has hardly been discussed in classic textbooks like \cite{pai} and \cite{kundur}.} Both of the multi-time-scale models in pp. $91$ of \cite{pai} and in pp. $86$ of \cite{kundur} use stator fluxes $\psi_d$ and $\psi_q$ as state variables, which does not lend itself easily to a current injection framework.

 Therefore, we consider stator currents $i_{sqd}$ ($= i_{sq} + ji_{sd}$) as state variables, combine the $R_g$-$L_g$ dynamics of the transformer with the $R_s$-$L_d^{''}$ dynamics of armature, and pose the Thevenin's equivalent voltage source behind the combined impedance as $E_{qd}$:

\vspace{-5pt}
\begin{equation}
\begin{array}{cc}
     E_{q} = E_{q}^{''} \omega_{g} + \frac{\dot{E}_{d}^{''}}{\omega_{s}} \hspace{30pt}
     E_{d} = E_{d}^{''} \omega_{g} - \frac{\dot{E}_{q}^{''}}{\omega_{s}}
\end{array}
\end{equation}
where, $E_{q}^{''}$ and $E_{d}^{''}$ are as in pp. $100$ of \cite{chaudhuri2014multi}.
As depicted in Fig.~\ref{fig:gen_ref_frame}, using the relationship $x_{DQ} = e^{j\delta_g} x_{q_g d_g}$, we transform all variables to the synchronously rotating $D$-$Q$ frame and in turn write the stator differential equation as 
\begin{equation} \label{eqn:gen_dyn_eqn}
    \begin{array}{l}
         \hspace{-10pt}\dot i_{sDQ} = \frac{\omega_{s}}{(L_{g}+L_{d}^{''})}[ E_{DQ} - v_{NDQ} -j \omega^*(L_{g}+L_{d}^{''}) i_{sDQ} \\ \hspace{25pt}- (R_{g}+R_{s}) i_{sDQ}].
    \end{array}
\end{equation}

\subsection{Modeling of GFC and its controls}
\begin{figure}[h]
    \centerline{
\includegraphics[width=0.37\textwidth]{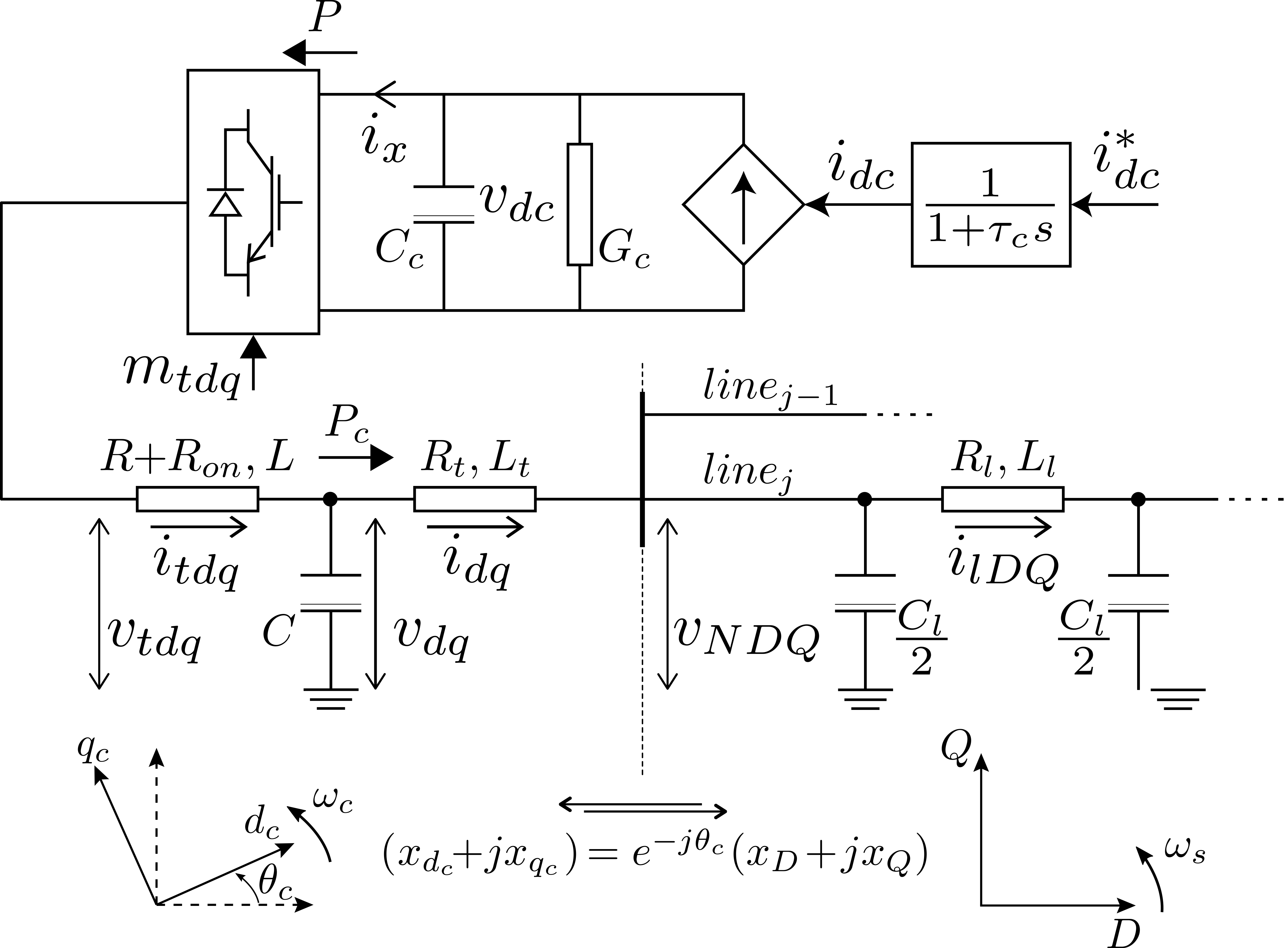}}
    \vspace{-5pt}
    \caption{Circuit model of GFC [parameters: $\tau_{c}$ = $0.05s$, $G_{c}$  = $0.45~pu$, $C_{c}$ = $325.73~pu$, $k_{dc}$ = $1080~pu$, $R$ = $5.5556e{-5}~pu$, $L$ = $0.0042~pu$, $C$ = $2.0358~pu$, $S_{base}$ = $100MVA$, $V_{dc,base}$ = $48.98kV$, $V_{ac,base}$ = $20kV$].}
    \label{fig:circuit_gfc}
\end{figure}

A generic circuit diagram of a GFC with a functional model of the renewable source is shown in Fig.~\ref{fig:circuit_gfc}, where reference current input $i_{dc}^*$ can be expresses as \cite{tayyebi}
$i_{dc}^* = k_{dc}(v_{dc}^*-v_{dc}) + G_c v_{dc} + \frac{P_{c}^*}{v_{dc}^*} + \frac{P_{c}^* - P_{t}}{v_{dc}^*}$  and converter ac terminal power $P_t = P$ assuming converter losses are added in the form of resistance $R_{on}$ to the filter resistance. 

\begin{figure}[h!]
    \centerline{
    \includegraphics[width=0.45\textwidth]{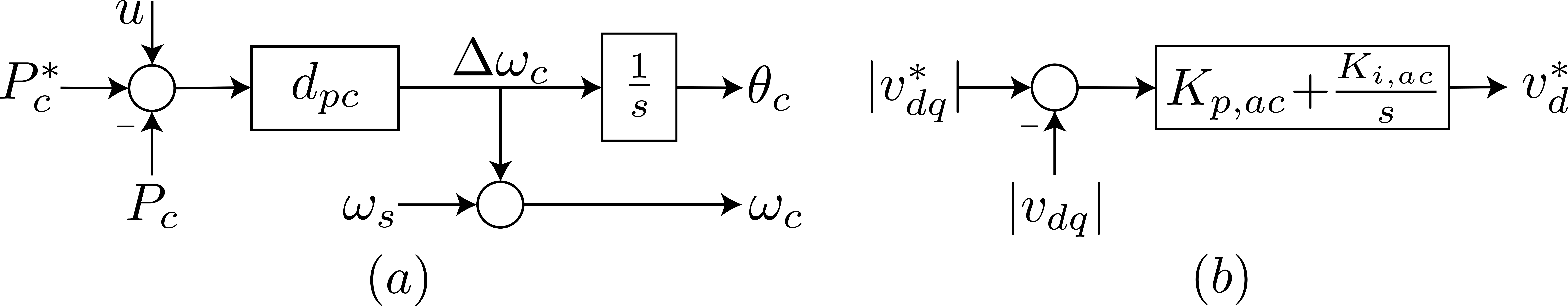}}
    \vspace{-5pt}
    \caption{(a) Power-frequency droop control, and (b) outer loop voltage control of GFCs [parameters: $\omega_{s}$ = $120\pi~rads^{-1}$, $P_{c}^*$ = $7.00~pu$, $K_{p,ac}$ = $0.0010$, $K_{i,ac}$ = $0.5000$, $S_{base}$ = $100MVA$].}
    \label{fig:Power-frequency droop control of GFC}
\end{figure}

\begin{figure}[h!]
    \centerline{
    \includegraphics[width=0.45\textwidth]{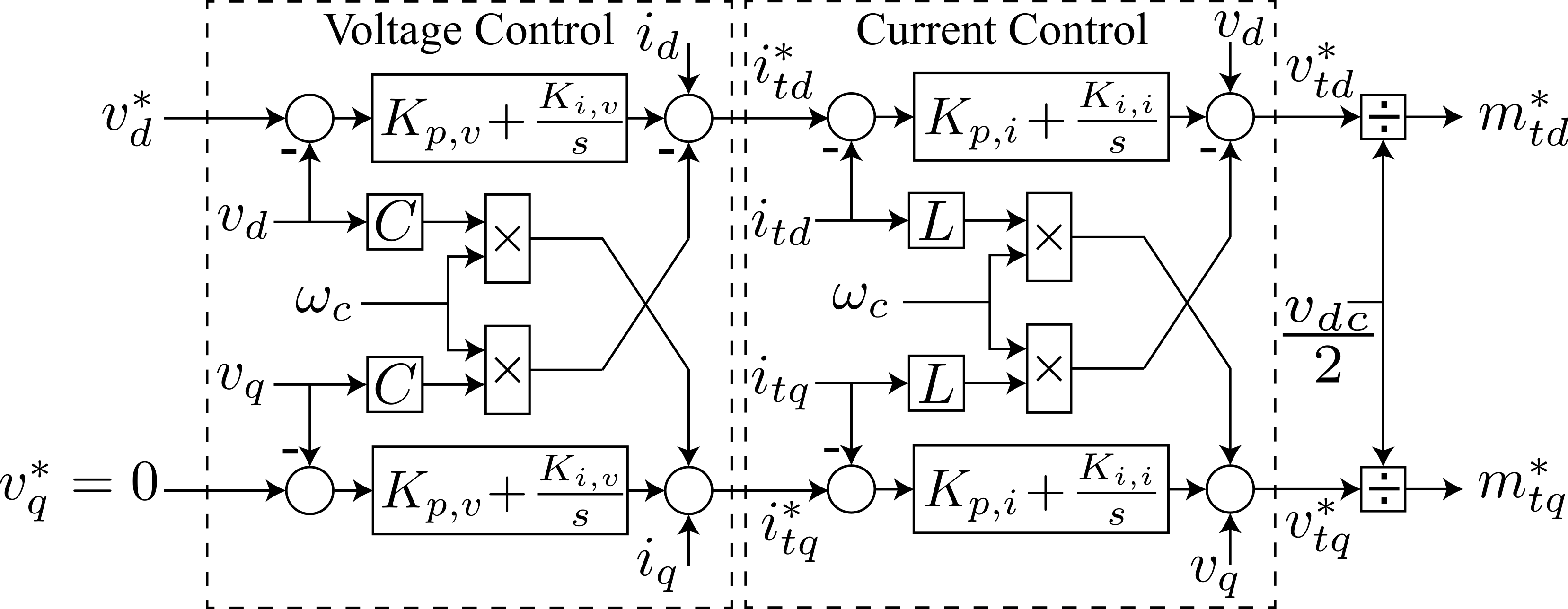}}
    \vspace{-5pt}
    \caption{Inner current and voltage control loops of GFC [parameters: $K_{p,i}$ = $0.0411~pu$, $K_{i,i}$ = $1.7389e{-4}~pu$, $K_{p,v}$ = $9.3600~pu$, $K_{i,v}$ = $0.0554~pu$, $S_{base}$ = $100MVA$, $V_{ac,base}$ = $20kV$].}
    \label{fig:Inner loop current and voltage controls of GFC}
\end{figure}

The ac side of the GFC is modeled and controlled in SPC framework that employs a rotating $d_c$-$q_c$ frame whose angular frequency $\omega_c$ is determined by the power frequency droop mechanism illustrated in Fig.~\ref{fig:Power-frequency droop control of GFC}(a). As shown in Fig.~\ref{fig:Inner loop current and voltage controls of GFC}, the standard inner current and voltage control loops are employed to regulate the current through the series $R-L$ component and the voltage across the capacitor $C$ in Fig.~\ref{fig:circuit_gfc}. The outer voltage control loop (Fig.~\ref{fig:Power-frequency droop control of GFC}(b)) regulates the voltage magnitude across the capacitor. Additionally, it makes sure that the voltage space-phasor at the PCC is aligned with the $d_c$-axis and maintains a zero $q_c$-axis voltage component.  \vspace{-8pt}
\vspace{-8pt}
\vspace{-8pt}

The ac-side dynamic model of the GFC considers the RLC filter, which connects the converter's ac terminal to the grid through a transformer that is represented by an equivalent series resistance $R_{t}$ and leakage inductance $L_{t}$. The ac-side model in $d_c$-$q_c$ frame uses state variable $v_{NDQ}$ from \eqref{eqn:con_dyn_eqn} as input (which is solved algebraically for QPC framework) after transforming it as shown in Fig.~\ref{fig:circuit_gfc}.
The KCL equation in~\eqref{eq:KCL_VL} is in current injection form where the injected current $i_{d_cq_c}$ is transformed into $D$-$Q$ frame and used as input to the transmission line model described earlier. The GFC model builds on a prior paper of the authors, see \cite{lilan_bakstep} for more details. \vspace{-6pt}


\section{Case Studies} \label{sec:case_study}
\begin{figure}[h!]
    \centerline{
    \includegraphics[width=0.41\textwidth]{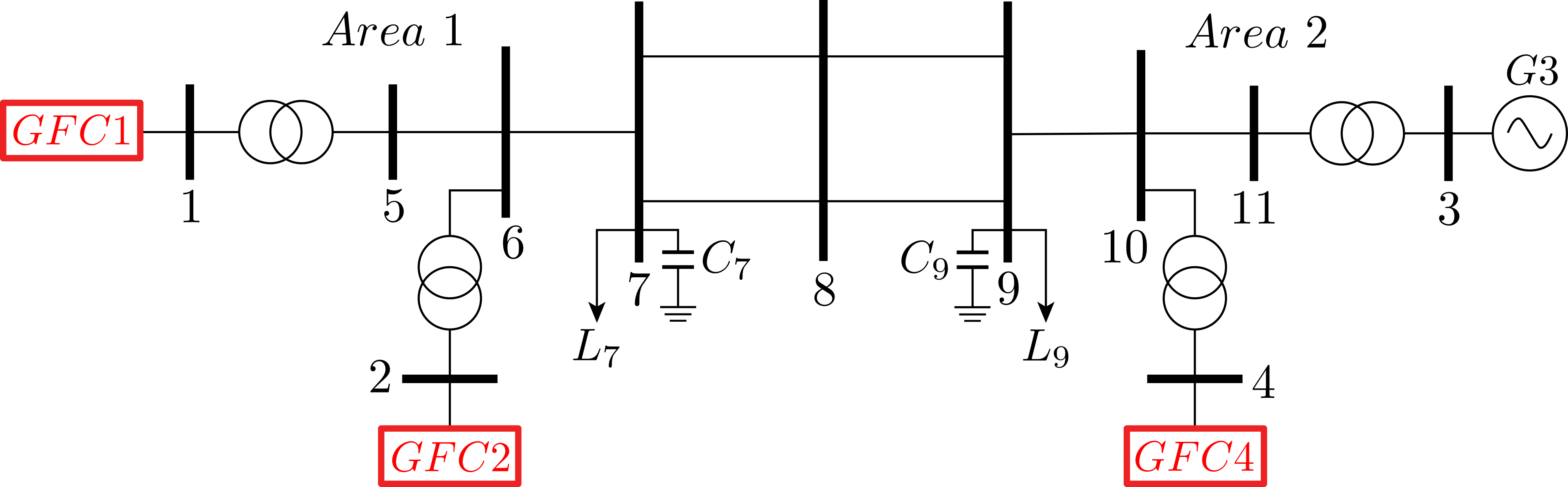}}
    \vspace{-5pt}
    \caption{Modified two-area test system with $1$-SG and $3$-GFCs.}
    \vspace{-2pt}
    \label{fig:11_bus_system}
\end{figure}

In this section we present case studies on the IEEE $4$-machine test system \cite{kundur} after replacing some or all of the SGs with GFCs while operating under nominal load flow condition with $400$ MW tie-flow. Due to space restrictions, we limit our analyses to two case studies: (1) when the modified system has $1$ SG and $3$ GFCs (see Fig.~\ref{fig:11_bus_system}) followed by (2) when $100\%$ IBR penetration is considered. The first case will be analyzed in detail whereas the second one will be discussed briefly. 
\vspace{-5pt}

To that end, QPC and SPC-based models of the test system shown in Fig.~\ref{fig:11_bus_system} are implemented in Matlab/Simulink \cite{simulink} environment \textit{in state-space form using the basic building blocks}. In addition, EMT model of the same system is built in EMTDC/PSCAD \cite{pscad}, which considers detailed switched models of the converters, distributed parameter representation of the transmission lines, and (static) constant impedance model of the loads. 

\vspace{-10pt}
\begin{table}[h!]
\label{tab:freq_unstable_mode}
\caption{Case (1): Unstable mode in different modeling approaches}
\centering
\begin{tabular}{ | l | *{3}{c|} }
    \hline
    \textbf{Framework} & \textbf{QPC} & \textbf{SPC} & \textbf{EMT} \\
    \hline
    Approach & Linearization & Linearization & Prony \\
    \hline
    $f, Hz$ & - & $43.135$ & $43.183$ \\
    \hline
    $\zeta, \%$ & - & $-2.11$ & $-3.40$ \\
    \hline
\end{tabular}
\end{table}

\begin{figure}[h!]
    \centerline{
    \includegraphics[width=0.45\textwidth]{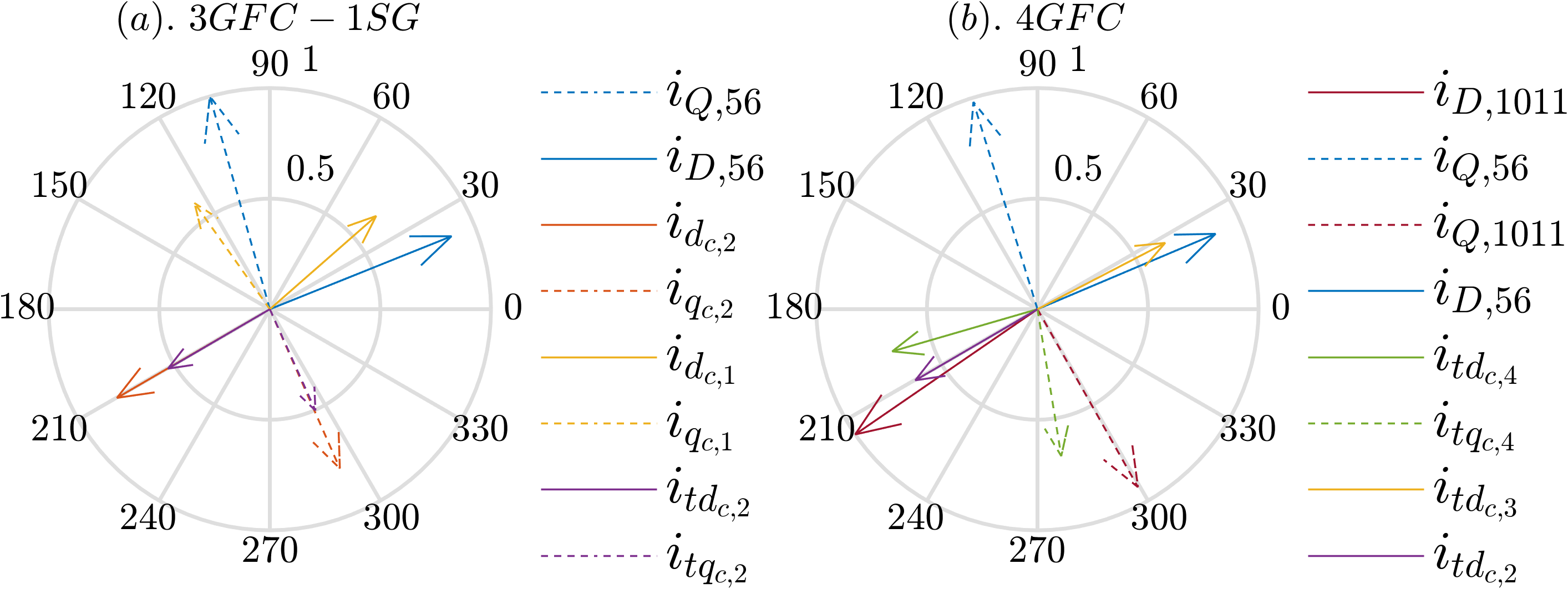}}
    \vspace{-8pt}
    \caption{SPC model: Compass plots of normalized participation factor magnitudes and modeshape angles of the dominant states contributing to the unstable eigenvalue pair. (a) Case (1), (b) Case (2).}
    \vspace{-15pt}
    \label{fig:mode_shapes}
\end{figure}

First, we perform frequency-domain analysis of linearized QPC and SPC-based models. It reveals a contradictory outcome, where the QPC model is stable with all eigenvalues in the left half of the $s$-plane while the SPC model exhibits an unstable SSO mode of $43.135$ Hz as shown in Table~I. Figure~\ref{fig:mode_shapes}(a) shows the compass plots of normalized participation factor magnitudes and modeshape angles of the dominant states contributing to the unstable $43$ Hz mode in the SPC model. 
The most dominant states are the current states of the transmission line $5$-$6$, which are absent in the QPC-based model and as a result it does not capture the unstable mode.

\begin{figure}[h!]
    \centerline{
    \includegraphics[width=0.44\textwidth]{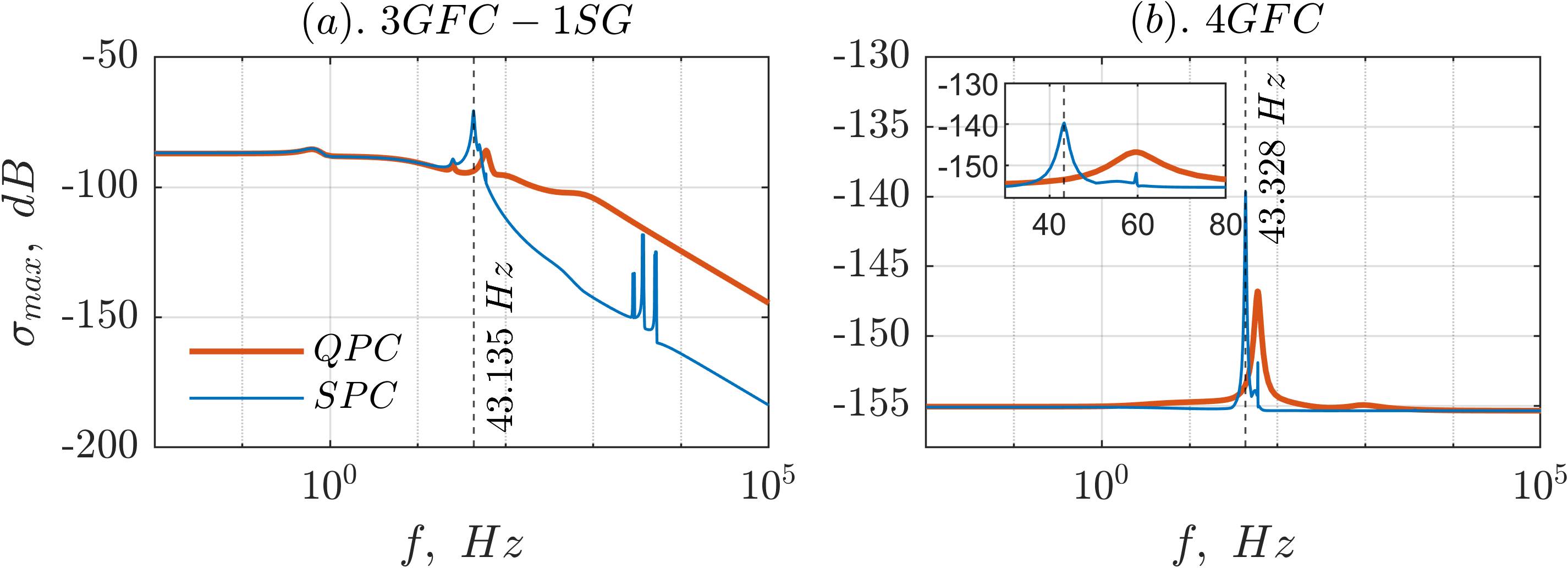}}
    \vspace{-8pt}
    \caption{Comparison of maximum singular values of QPC and SPC models. (a) Case (1), (b) Case (2).}
    \vspace{-4pt}
    \label{fig:sv_plot}
\end{figure}

The maximum singular value plots of the $3$-input-$9$-output linearized models based on QPC and SPC are compared in Fig.~\ref{fig:sv_plot}(a), where signal $u$ modulating power references to the GFCs (see Fig.~\ref{fig:Power-frequency droop control of GFC}(a)) are inputs, and the dc-link voltages and the currents of transformers of the GFCs are outputs. Besides the spikes in the spectrum at higher frequencies due to the network modes, a sharp peak in gain can be observed at the frequency of $43$ Hz for SPC-based model, which are absent in the QPC-based model. \vspace{-5pt}

\begin{figure}[h!]
    \centerline{
    \includegraphics[width=0.4\textwidth]{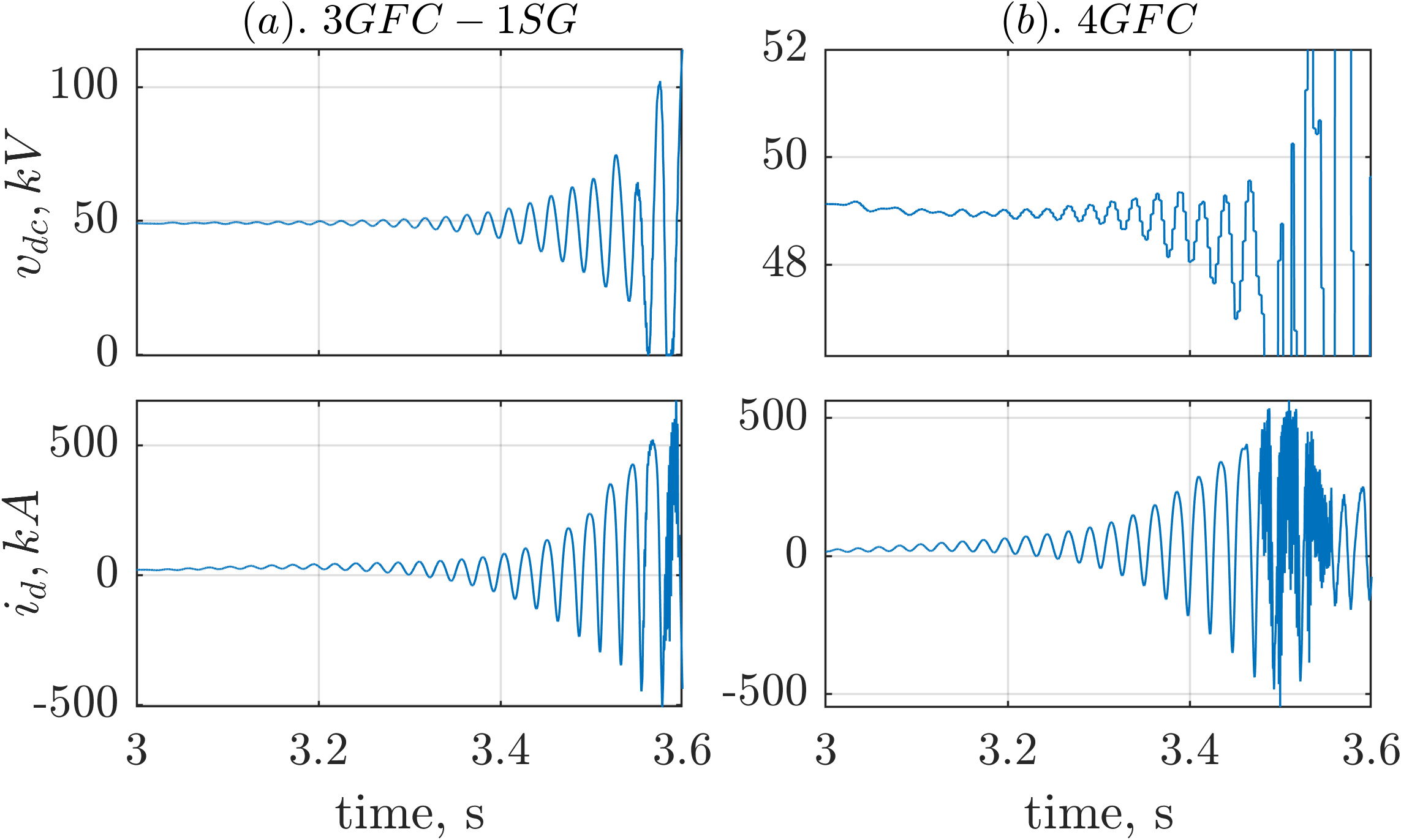}}
    \vspace{-8pt}
    \caption{GFC$1$ responses of dc-link voltage and $d$-axis current at the PCC obtained from EMT simulations. (a) Case (1), (b) Case (2).}
    \vspace{-4pt}
    \label{fig:EMP_time_domain_sim}
\end{figure}

Figure~\ref{fig:EMP_time_domain_sim}(a) indicates the time-domain response of dc-link voltage and $d$-axis current of the transformer of GFC$1$ from EMT simulation. Prony analysis \cite{prony} on this response reveals that the frequency of oscillations is $43$ Hz with -$3.4$\% damping ratio, which is pretty close to what we observe in the eigenvalue analysis of SPC model, see Table~I. \vspace{-5pt}


Finally, for case (2) with $4$ GFCs the normalized participation factors corresponding to the unstable $43.33$ Hz mode is shown in Figs~\ref{fig:mode_shapes}(b). Maximum singular value plots (Fig.~\ref{fig:sv_plot}(b)) indicate that it is captured by the SPC model, whereas QPC model does not capture it. Time-domain responses from the EMT model are shown in Fig. \ref{fig:EMP_time_domain_sim}(b). These findings are in line with those of case (1).

\section{Limitations \& Conclusions}\label{sec:Conclusion}
The SPC-based model in $d$-$q$ frame that considers 
lumped parameter transmission network dynamics is shown to capture unstable $43$ Hz modes in presence of GFCs, which do not show up in the QPC model. The caveat is that the accuracy of such results depend upon the degree of approximation involved in the lumped-parameter $\pi$-model with respect to the actual distributed transmission line model. Notwithstanding the advantages of SPC-based models compared to EMT models (e.g., faster simulation,  scalability, linearizability), extreme caution should be exercised before drawing firm conclusions that involve lumped representation of network dynamics. Future research should be focused on more accurate transmission line models for GFC-dominated systems.   

\bibliographystyle{IEEEtran}
\bibliography{reference.bib}

\begin{thebibliography}{10}
\providecommand{\url}[1]{#1}
\csname url@samestyle\endcsname
\providecommand{\newblock}{\relax}
\providecommand{\bibinfo}[2]{#2}
\providecommand{\BIBentrySTDinterwordspacing}{\spaceskip=0pt\relax}
\providecommand{\BIBentryALTinterwordstretchfactor}{4}
\providecommand{\BIBentryALTinterwordspacing}{\spaceskip=\fontdimen2\font plus
\BIBentryALTinterwordstretchfactor\fontdimen3\font minus \fontdimen4\font\relax}
\providecommand{\BIBforeignlanguage}[2]{{%
\expandafter\ifx\csname l@#1\endcsname\relax
\typeout{** WARNING: IEEEtran.bst: No hyphenation pattern has been}%
\typeout{** loaded for the language `#1'. Using the pattern for}%
\typeout{** the default language instead.}%
\else
\language=\csname l@#1\endcsname
\fi
#2}}
\providecommand{\BIBdecl}{\relax}
\BIBdecl

\bibitem{IBRSSOTF}
Y.~Cheng, L.~Fan, J.~Rose, S.-H. Huang, J.~Schmall, X.~Wang, X.~Xie, J.~Shair, J.~R. Ramamurthy, N.~Modi, C.~Li, C.~Wang, S.~Shah, B.~Pal, Z.~Miao, A.~Isaacs, J.~Mahseredjian, and J.~Zhou, ``Real-world subsynchronous oscillation events in power grids with high penetrations of inverter-based resources,'' \emph{IEEE Transactions on Power Systems}, vol.~38, no.~1, pp. 316--330, 2023.

\bibitem{Lingling-InterIBR}
L.~Fan, ``Inter-{IBR} oscillation modes,'' \emph{IEEE Transactions on Power Systems}, vol.~37, no.~1, pp. 824--827, 2022.

\bibitem{Strunz-21-QPC-DPC}
J.~Vega-Herrera, C.~Rahmann, F.~Valencia, and K.~Strunz, ``Analysis and application of quasi-static and dynamic phasor calculus for stability assessment of integrated power electric and electronic systems,'' \emph{IEEE Transactions on Power Systems}, vol.~36, no.~3, pp. 1750--1760, 2021.

\bibitem{Mani-94-DynamicPhasor}
V.~Venkatasubramanian, ``Tools for dynamic analysis of the general large power system using time-varying phasors,'' \emph{International Journal of Electrical Power \& Energy Systems}, vol.~16, no.~6, pp. 365--376, 1994.

\bibitem{Verghese-91-DynamicPhasor}
S.~Sanders, J.~Noworolski, X.~Liu, and G.~Verghese, ``Generalized averaging method for power conversion circuits,'' \emph{IEEE Transactions on Power Electronics}, vol.~6, no.~2, pp. 251--259, 1991.

\bibitem{Stankovic-00-DynamicPhasor}
A.~Stankovic and T.~Aydin, ``Analysis of asymmetrical faults in power systems using dynamic phasors,'' \emph{IEEE Transactions on Power Systems}, vol.~15, no.~3, pp. 1062--1068, 2000.

\bibitem{Bozhko-16-DynamicPhasor}
T.~Yang, S.~Bozhko, J.-M. Le-Peuvedic, G.~Asher, and C.~I. Hill, ``Dynamic phasor modeling of multi-generator variable frequency electrical power systems,'' \emph{IEEE Transactions on Power Systems}, vol.~31, no.~1, pp. 563--571, 2016.

\bibitem{Mani-95-DQ0}
V.~Venkatasubramanian, H.~Schattler, and J.~Zaborszky, ``Fast time-varying phasor analysis in the balanced three-phase large electric power system,'' \emph{IEEE Transactions on Automatic Control}, vol.~40, no.~11, pp. 1975--1982, 1995.

\bibitem{Demiray2008}
T.~H. Demiray, ``Simulation of power system dynamics using dynamic phasor models,'' Ph.D. dissertation, ETH Zurich, 2008.

\bibitem{Stankovic-02-DP}
A.~Stankovic, S.~Sanders, and T.~Aydin, ``Dynamic phasors in modeling and analysis of unbalanced polyphase ac machines,'' \emph{IEEE Transactions on Energy Conversion}, vol.~17, no.~1, pp. 107--113, 2002.

\bibitem{pai}
P.~W. Sauer and M.~A. Pai, \emph{Power System Dynamics and Stability}.\hskip 1em plus 0.5em minus 0.4em\relax Prentice Hall, 1998.

\bibitem{chaudhuri2014multi}
N.~Chaudhuri, B.~Chaudhuri, R.~Majumder, and A.~Yazdani, \emph{Multi-terminal Direct-Current Grids: Modeling, Analysis, and Control}, ser. IEEE Press.\hskip 1em plus 0.5em minus 0.4em\relax Wiley, 2014.

\bibitem{kundur}
P.~Kundur, \emph{Power System Stability and Control}.\hskip 1em plus 0.5em minus 0.4em\relax McGraw-Hill, 1994.

\bibitem{tayyebi}
A.~Tayyebi, D.~Groß, A.~Anta, F.~Kupzog, and F.~Dörfler, ``Frequency stability of synchronous machines and grid-forming power converters,'' \emph{IEEE Journal of Emerging and Selected Topics in Power Electronics}, vol.~8, no.~2, pp. 1004--1018, 2020.

\bibitem{lilan_bakstep}
L.~Karunaratne, N.~R. Chaudhuri, A.~Yogarathnam, and M.~Yue, ``Nonlinear backstepping control of grid-forming converters in presence of grid-following converters and synchronous generators,'' \emph{IEEE Transactions on Power Systems}, vol.~39, no.~1, pp. 1948--1964, 2024.

\bibitem{simulink}
\emph{{MATLAB version 9.8.0.1396136 (R2020a)}}, The Mathworks, Inc., Natick, Massachusetts, 2020.

\bibitem{pscad}
\emph{{PSCAD/EMTDC User’s Manual}}, Manitoba HVDC Research Centre, Winnipeg, Canada, 1998.

\bibitem{prony}
G.~Liu, J.~Quintero, and V.~M. Venkatasubramanian, ``Oscillation monitoring system based on wide area synchrophasors in power systems,'' in \emph{2007 iREP Symposium - Bulk Power System Dynamics and Control - VII. Revitalizing Operational Reliability}, 2007, pp. 1--13.

\end{thebibliography}


\end{document}